\documentclass[a4paper,oneside, onecolumn,final,times,number,3p]{elsarticle}
\usepackage{amsmath,amsthm}
\usepackage{mathtools}
\usepackage{framed}
\usepackage{multicol} 
\usepackage{nomencl} 

\makenomenclature

\renewcommand*\nompreamble{\begin{multicols}{2}}

\renewcommand*\nompostamble{\end{multicols}}
\usepackage{graphics}

\usepackage{epsfig}
\usepackage{float}
\usepackage{epstopdf}
\usepackage{enumitem}
\graphicspath{{./images/}}

\usepackage{graphicx}
\usepackage{caption}
\usepackage{subcaption}

\makeatletter
\g@addto@macro\@floatboxreset\centering
\makeatother

\makeatletter
\@fpsep\textheight
\makeatother
\extrafloats{100}
\usepackage{caption}
\journal{   }

\begin{document}
\begin{frontmatter}

\title {Numerical simulations of fluid flow and heat transfer in a four-sided lid-driven rectangular domain}

\author[rvt]{V. Ambethkar\corref{cor1}}
\ead{vambethkar@maths.du.ac.in, vambethkar@gmail.com}
\author[rvt]{Durgesh Kushawaha}
\ead{dkushawaha@maths.du.ac.in, durgeshoct@gmail.com}

\cortext[cor1]{Corresponding author}

\address[rvt]{Department of Mathematics,\\
Faculty of Mathematical Sciences,\\
University of Delhi\\ Delhi,
110007, India}

\begin{abstract}
Numerical simulations for 2-D unsteady, incompressible flow with heat transfer in a four-sided lid-driven rectangular domain are reported in the present study. For the four-sided lid-driven rectangular domain, the lower wall is moved to the left, the upper wall is moved to the right, while the right wall is moved upwards and the left wall is moved downwards. All four walls move with equal speed. Different constant temperatures are applied to the left and right moving walls, and thermal insulation is applied to the upper and bottom moving walls. The governing equations are discretized using the QUICK scheme of finite volume methods. The SIMPLE algorithm is adopted to compute the numerical solutions of the flow variables, $u$-velocity, $v$-velocity, $P$, and $\theta$ as well as local and average Nusselt numbers for $50 \le Re \le 1500$   and $Pr=6.63$. Due to the force generated by moving fluid, the direction of moving walls and the Reynolds number affect fluid flow in the rectangular domain in addition, at different Reynolds numbers along the cold wall of the domain, the variation in average and local Nusselt numbers reveals that overall heat transfer increases isotherms showed that as Reynolds numbers increase, the horizontal temperature gradient near the vertical walls decreases, because of which heat transfer decreases.
\end{abstract}

\end{frontmatter}

\begin{table*}[!t]

\begin{framed}

\printnomenclature[8.0mm]

\end{framed}
\end{table*}

\section{Introduction}
\label{sec1}
Fluid flow and heat transfer in a four-sided lid-driven rectangular domain has been the subject of intensive research in recent years. This is due to its significant applications such as cooling of electronic devices, furnaces, heat exchangers, boiler tubes, cooling of cylinder heads in I.C. engines, heating and cooling of buildings, heating of electric irons, heat treatment of engineering components, quenching of ingots, freezing of foods, etc.
     
Extensive literature studies have focused on heat transfer and fluid flow in rectangular or square cavities. These studies fall into two categories. The first deals with horizontal top \cite{b1}-\cite{b10} or bottom \cite{b11} wall sliding lid-driven two-dimensional cavities, in which the top wall has a constant velocity \cite{b1}-\cite{b3} or oscillates \cite{b4,b5}, and behaves similarly in three dimensional cavities \cite{b6,b7,b9,b10,b16}. Various boundary conditions are applied to other solid walls in such cases. The second one is concerned with side-driven, differentially heated cavities in these cases, left or right vertical wall or both vertical walls move with a constant velocity in their planes \cite{b12,b13,b14} in these studies, to create a temperature gradient in the domain usually the lid-driven side and the one opposing are heated differentially.
     
Arpaci and Larsen \cite{b12} presented an analysis of the mixed-convection heat transfer in tall cavities in their study, one vertical side moved, vertical boundaries were maintained at different temperatures, and the horizontal boundaries were adiabatic.
     
Aydn \cite{b13} numerically studied mechanisms of aiding and opposing forces in a shear- and buoyancy-driven cavity. The square cavity had one vertical hot wall moving upwards or downwards, the opposite cold wall fixed, and both horizontal walls adiabatic. Oztop and Dagtekin \cite{b14} examined mixed convection in a two-sided, lid-driven differentially heated square cavity.

Kuhlmann et al. \cite{b15} conducted a numerical and experimental study on steady flow in rectangular two-sided lid-driven cavities. They found that the basic two-dimensional flow was not always unique. For low Reynolds numbers, it consists of two separate co-rotating vortices adjacent to the moving walls.

Blohm and Kuhlmann \cite{b16} studied experimentally incompressible fluid flow in a rectangular container driven by two facing side walls which move steadily in anti-parallel for Reynolds numbers up to 1200. Two rotating cylinders of large radii close the cavity  tightly, and create the moving side walls. Beyond a first threshold, three-dimensional cells bifurcate supercritically out of the basic flow state. When both side walls move at the same rate (driven symmetrically), oscillatory stability was tricritical.

Xu et al. \cite{b17} have investigated the unsteady flow with heat transfer adjacent to the finned side wall of a differentially heated cavity with conducting adiabatic fin. Basak et at. \cite{b18} have investigated the effects of thermal boundary conditions on natural convection flows within a square cavity. Ilis et at. \cite{b19} have studied the effect of aspect ratio on entropy generation in a rectangular cavity with differentially heated vertical walls. Wahba \cite{b21} has investigated the multiplicity of states for two-sided and four-sided lid driven cavity flows.      
     
The above-mentioned literature survey pertinent to the present problem under consideration revealed that lid-driven cavities have interesting applications in various fields. However, no studies in the literature considered the case of a four-sided lid driven rectangular domain with fluid flow and heat transfer. Specifically, in the present paper discussing the four-sided driven rectangle, the lower wall is moved to the left, the upper wall is moved to the right, while the right wall is moved upwards and the left wall is moved downwards. All four walls move with equal speed.

What motivated us is the enormous scope of applications of unsteady incompressible flow with heat transfer as discussed earlier. Literature survey also revealed that the problem of fluid flow and heat transfer in a four-sided lid-driven rectangular domain, along with slip wall, and temperature boundary conditions, has not been studied numerically. Furthermore, to investigate the importance of the applications enumerated earlier, there is a need to determine numerical solutions of the unknown flow variables to fulfil this requirement, we present numerical simulations of the problem of fluid flow and heat transfer in a four-sided lid-driven rectangular domain, along with slip wall and temperature boundary conditions, using the SIMPLE algorithm.

Our main target of this work is to numerically investigate fluid flow and heat transfer in a four-sided lid-driven rectangular domain. We are used the QUICK scheme of finite volume methods to discretize the governing equations. SIMPLE algorithm is adopted to compute the numerical solutions of the flow variables, $u$-velocity, $v$-velocity, $P$, and $\theta$ as well as local and average Nusselt numbers for $50 \le Re \le 1500$ and $Pr=6.63$ at time $t=0.001$s.
     
The summary of the layout of the current work is as follows: Section 2 describes mathematical formulation that includes the physical description of the problem, governing equations, and initial and boundary conditions. Section 3 describes the numerical solution of the governing equations along with validity of results obtain with the benchmark solutions. Section 4 discusses the numerical results. Section 5 illustrates the conclusions of this study.

\section{Mathematical Formulation}
\label{sec2}
\subsection{Physical Description}
\label{sec2.1.}

\begin{figure}[H]
\includegraphics[scale=0.8]{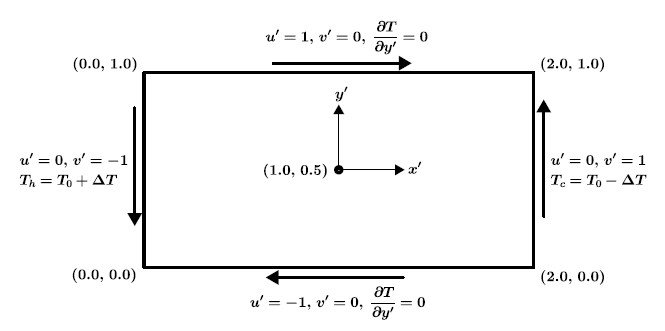}
\caption{Geometry of the four-sided lid driven rectangular domain problem.}
\label{f1}
\end{figure}

Geometry of the problem considered in this work along with the boundary conditions is depicted in the Fig.~\ref{f1} A four-sided lid-driven rectangular domain around the point $(1,0.5)$ in which laminar unsteady incompressible flow is considered. The lower wall is moved to the left, the upper wall is moved to the right, while the right wall is moved upwards and the left wall is moved downwards. All four walls move with equal speed. The vertical lids have different constant temperatures. The horizontal walls are adiabatic. The left wall considered as the hot wall and the right is as the cold wall.

We are assuming that, the values of $T_0$ and $\Delta T$ are chosen suitably so that the temperature defined on the left wall is greater than that of right wall. We are assumed that, at all four corner points of the computational domain, velocity components $\left(u,\,v \right )$ vanish. It may be noted here regarding specifying the boundary conditions for pressure, the convention followed is that either the pressure at boundary is given or velocity components normal to the boundary are specified \cite{b20}.

\subsection{Governing equations}
\label{sec2.2.}

The flow is assumed to be two-dimensional, unsteady state, laminar, and the fluid is incompressible. The dimensionless forms of the governing equations are the continuity, and the $x$- and $y$-components of the Navier-Stokes and the energy equations, assuming negligible dissipation and constant thermo-physical properties, as given below:

\begin{align}
&\mbox{Continuity equation}
&&
\dfrac{\partial u}{\partial x}+\dfrac{\partial v}{\partial y}=0,
\label{e1}
\\[2mm]
&\mbox{$x$-momentum}
&&
\dfrac{\partial u}{\partial t} +u\dfrac{\partial u}{\partial x}+v\dfrac{\partial u}{\partial y}=-
\dfrac{\partial P}{\partial x}+
\left(\dfrac{1}{Re}\right) \left(
\dfrac{\partial^2 u}{\partial x^2}+
\dfrac{\partial^2 u}{\partial y^2}\right),
\label{e2}
\\[2mm]
&\mbox{$y$-momentum}
&&
\dfrac{\partial v}{\partial t}+
u\dfrac{\partial v}{\partial x}+
v\dfrac{\partial v}{\partial y}=-
\dfrac{\partial P}{\partial y}+
\left(\dfrac{1}{Re}\right) \left(
\dfrac{\partial^2 v}{\partial x^2}+
\dfrac{\partial^2 v}{\partial y^2}\right),
\label{e3}
\\[2mm]
&\mbox{Energy equation}
&&
\dfrac{\partial \theta}{\partial t}+
u\dfrac{\partial \theta}{\partial x}+
v\dfrac{\partial \theta}{\partial y}=\left(\dfrac{1}{Pr}\right)
\left(\dfrac{\partial^2 \theta}{\partial x^2}+
\dfrac{\partial^2 \theta}{\partial y^2}\right).
\label{e4}
\end{align}
where $u$, $v$, $P$, $\theta$, $Re$, and $Pr$ are the dimensionless velocity components in $x$- and $y$-directions, the dimensionless pressure, the dimensionless temperature, the Reynolds number, and the Prandtl number respectively.\medskip

We define the following non-dimensional variables

\begin{align}
\left.\begin{matrix}
x=\frac{{x}'}{L}, &y=\frac{{y}'}{H},  &v=\frac{{v}'}{V_p},  &u=\frac{{u}'}{V_p}, \\[2mm] 
P=\frac{{p}'}{\rho V_{p}{^2}},&\theta=\frac{T-T_c}{T_h-T_c},  &Re=\frac{V_pL}{\nu},  &Pr=\frac{\nu}{\alpha}. 
\end{matrix}\right\}\hspace{5.5 cm}
\label{e5}
\end{align}

The initial, no-slip and slip wall boundary conditions are given by:
\begin{align}
&\text{for $t= 0$,}~~
u(x,y,0)=0,\quad v(x,y,0)=0,\quad \theta(x,y,0)=295.55.
\label{e6} \hspace{2.5 cm}
\\[2mm]
&\!\!\!\!\left.
\begin{array}{lll}
\text{for $t>0$,}&
\text{$x=0$,\quad $0<y<1$,\quad $u=0$,\quad $v=-1$,\quad $\theta_{h}=\theta_{0}+\Delta{\theta}$,}
\\[2mm]
&\text{$x=2$,\quad $0<y<1$,\quad $u=0$,\quad $v=1$,\quad $\theta_{c}=\theta_{0}-\Delta{\theta}$,}
\\[2mm]
&\text{$y=0$,\quad $0<x<2$,\quad $u=-1$,\quad $v=0$,\quad $\dfrac{\partial \theta}{\partial y}=0$,}
\\[4mm]
&\text{$y=1$,\quad $0<x<2$,\quad $u=1$,\quad $v=0$,\quad $\dfrac{\partial \theta}{\partial y}=0$.}
\end{array}\right\} \hspace{2.5 cm}
\label{e7}
\end{align}

Local Nusselt number is 

\begin{align}
Nu_y=-\frac{1}{A}\frac{\left ( \frac{\partial \theta}{\partial x} \right )_w}{\left(\theta_h-\theta_c \right )}.\hspace{9 cm}
\end{align}
Integration of the local Nusselt number along the wall is used to calculate the average Nusselt number.
\begin{align}
\overline{Nu}=\frac{1}{A}\int_{0}^{1}Nu_y dx.\hspace{9 cm}
\end{align}

The stream function is calculated from the definition 
\begin{align}
u=\frac{\partial \psi}{\partial y},\quad v=-\frac{\partial \psi}{\partial x}.\hspace{9 cm}
\end{align}
It is taken $\psi=0$ at the solid boundaries.

\section{Validation of the numerical solutions}
\label{sec3.}

We are used a finite volume method to solve the governing equations in the discretizing controlling equations, the second order Quick scheme is selected. SIMPLE algorithm is adopted in solving the conservation equations. Discretization of the governing equations using Quick scheme is being skipped here as it is available in the literature \cite{b20,b22}.

All the results obtained in this study converged to a maximum residual of $10^{-8}$. Furthermore, we are considered three different grid systems: $(100\times 50)$, $(150\times 75)$, and $(200\times 100)$ to ensure grid-independent results and $(150\times 75$) grid points were considered to obtain accurate solution in the entire computation of this study. We are also noted that by increasing the grid density by $77\%$, from $(150 \times 75)$ to $(200\times 100)$, the relative change in the value of the average Nusselt number is less than $1\%\,(0.43\%)$, which confirm that computed results on the $(200\times 100)$ grid are indeed grid-independent.

{\renewcommand{\arraystretch}{1.5}
		\tabcolsep=1.0\tabcolsep
		\begin{table}[htb]\small
			\begin{center}
				\begin{tabular}{llll}
				\hline 
$Re$ number & Iwatsu et al. \cite{b2} $\overline{Nu}$ & Oztop and Dagtekin \cite{b14} $\overline{Nu}$ & This study $\overline{Nu}$\tabularnewline
\hline 
100 & 1.34 & 1.33 & 1.30\tabularnewline
400 & 3.62 & 3.6 & 3.59\tabularnewline
1000 & 6.29 & 6.21 & 6.21\tabularnewline
\hline 
				
				\end{tabular}
				\caption{Comparison of the average Nusselt number obtained in the present study with Iwatsu et al. \cite{b2} and Oztop and Dagtekin \cite{b14}.}
				\label{t1}
			\end{center}
		\end{table}
}	

The validation of current simulation has been verified with Iwatsu et al. \cite{b2} and Oztop and Dagtekin \cite{b14}. There is a good agreement for the average Nusselt numbers in the current study when compared to those of \cite{b2} and \cite{b14} as shown in table 1.

\section{Results and Discussion}
\label{sec4.}

Fluid flow and temperature fields in a four-sided lid-driven rectangular domain are examined. The numerical simulations are performed with the same Reynolds numbers on both sides of the domain. Simulations in the rectangular domain are performed for the Reynolds number range from 50 to 1500. The Four-sided lid-driven rectangular domain is analysed according to the direction of the moving plate shown in Fig.~\ref{f1}

\begin{figure}[H]
\centering
\begin{subfigure}{.5\textwidth}
  \centering
  \includegraphics[scale=0.4]{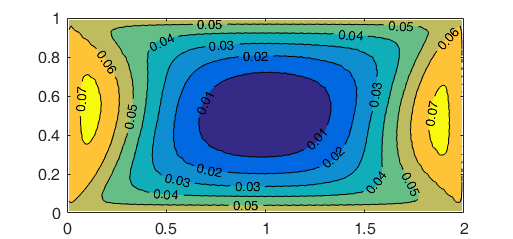}
  \caption{}
\end{subfigure}%
\begin{subfigure}{.4\textwidth}
  \centering
  \includegraphics[scale=0.4]{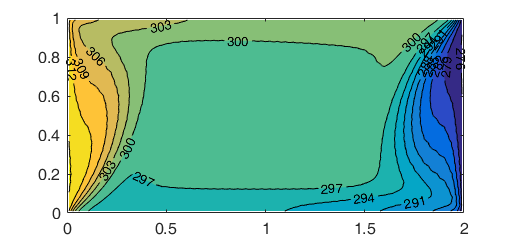}
 \caption{}
\end{subfigure}
\begin{subfigure}{.5\textwidth}
  \centering
  \includegraphics[scale=0.4]{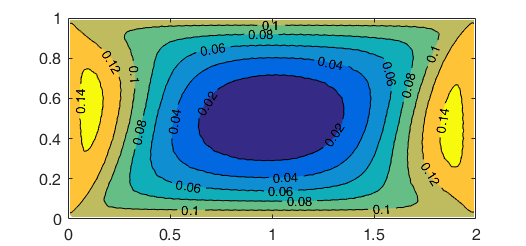}
 \caption{}
\end{subfigure}%
\begin{subfigure}{.4\textwidth}
  \centering
  \includegraphics[scale=0.4]{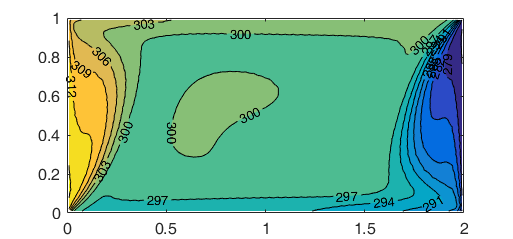}
  \caption{}
\end{subfigure}
\caption{Streamlines $(a-c)$ and isotherms $(b-d)$ for: $Re=50$, and $Re=100$.}
\label{f2}
\end{figure}

Streamlines $(a-c)$ and isotherms $(b-d)$ for $Re=50$ and 100 are shown in Fig.~\ref{f2} For $Re=50$ and $Re=100$ in Fig. 2(a) and (c), we observe, there is a primary cell with two secondary cells: the primary cell is at the center of the domain, however, the cell is not quite centered on the symmetry lines. The two secondary cells are weak and formed near the moving walls on both sides.

For Fig. 2(a) and (c), $\left | \psi_{ext}  \right |=7.46 \times 10^{-6}$, $x=1.01,\,y=0.51$ and $\left | \psi_{ext}  \right |=1.28 \times 10^{-5}$, $x=1.01,\,y=0.51$ respectively.

 \begin{figure}[H]
\centering
\begin{subfigure}{.5\textwidth}
  \centering
  \includegraphics[scale=0.4]{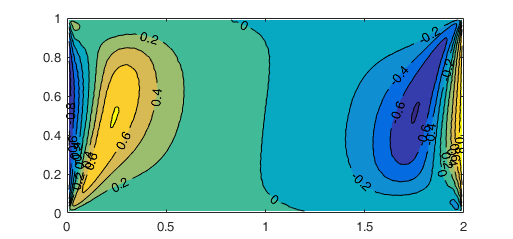}
  \caption{}
\end{subfigure}%
\begin{subfigure}{.4\textwidth}
  \centering
  \includegraphics[scale=0.4]{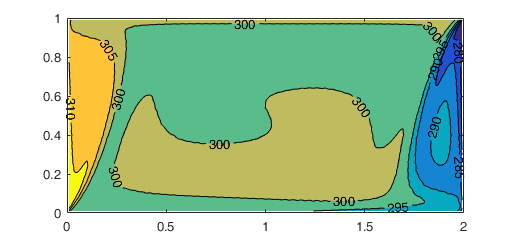}
 \caption{}
\end{subfigure}

 \caption{$(a)$ Streamlines and $(b)$isotherms for: $Re=500$.}
  \label{f3}
 \end{figure}

Streamlines and isotherms for $Re=500$ shown in Fig.~\ref{f3} For $Re=500$ in Fig. 3(a), we observe that there are two secondary cells only. Both weaker cells are near the moving left and right walls. We also observe that streamlines become dense near the moving left and right walls 

For Fig. 3(a), $\left | \psi_{ext}  \right |=-1.68 \times 10^{-3}$, $x=1.01,\,y=0.51$.

 \begin{figure}[H]
\centering
\begin{subfigure}{.5\textwidth}
  \centering
  \includegraphics[scale=0.4]{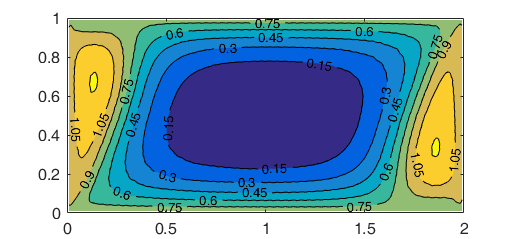}
  \caption{}
\end{subfigure}%
\begin{subfigure}{.4\textwidth}
  \centering
  \includegraphics[scale=0.4]{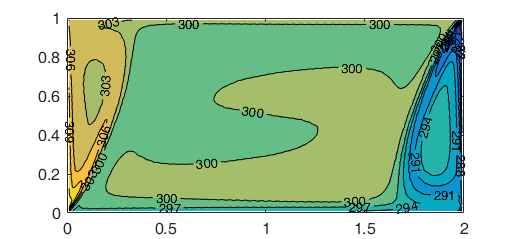}
 \caption{}
\end{subfigure}
\begin{subfigure}{.5\textwidth}
  \centering
  \includegraphics[scale=0.4]{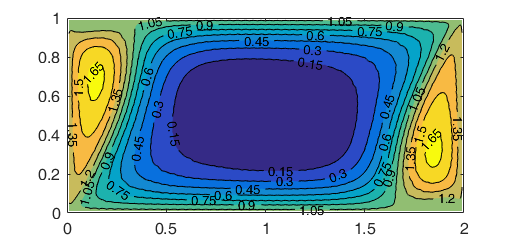}
 \caption{}
\end{subfigure}%
\begin{subfigure}{.4\textwidth}
  \centering
  \includegraphics[scale=0.4]{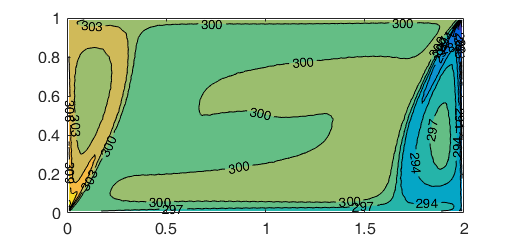}
  \caption{}
\end{subfigure}
\caption{Streamlines $(a-c)$ and isotherms $(b-d)$ for: $Re=1000$, and $Re=1500$.}
  \label{f4}
 \end{figure}

Streamlines $(a-c)$ and isotherms $(b-d)$ for $Re=1000$ and 1500 are shown in Fig.~\ref{f4} For $Re=1000$ and $Re=1500$ in Fig. 4(a) and (c), we observe, there is a primary cell with two secondary cells: the primary cell is at the center of the domain, however, the cell is not quite centered on the symmetry lines. The two secondary cells are weak and formed near the moving walls on both sides.

For Fig. 4(a) and (c), $\left | \psi_{ext}  \right |=2.54 \times 10^{-5}$, $x=1.01$, $y=0.51$ and $\left | \psi_{ext}  \right |=2.36 \times 10^{-5}$, $x=1.01$, $y=0.51$ respectively.  
 
From Fig. 2(a) and (c), Fig. 4(a) and (c), we observe, there is a primary cell with two secondary cells: the primary cell is at the center of the domain; however, the cell is not quite centered on the symmetry lines. The two secondary cells are weak and formed near the moving walls on both sides. From Fig. 3(a), we observe that there are two secondary cells only. Both weaker cells are near the moving left and right walls. We also observe that streamlines become dense near the moving left and right walls.
     
This happens because of the fluid rises along the right cold wall and sinks on the left hot wall due to forces generated by the moving fluid. Formation of two rotating cells at each side and a rotating cell at the center makes the heat transfer from left to right possible. The same phenomena have not been observed with one and two vertical- sided lid-driven cavities in the literature isotherms in Fig. 2(b), (d), Fig. 3(b), and Fig. 4(b) and (d) show that as the Reynolds numbers increase the horizontal temperature gradient near left and right vertical walls decreases, because of which heat transfer decreases. 

 \begin{figure}[H]
  \includegraphics[scale=0.8]{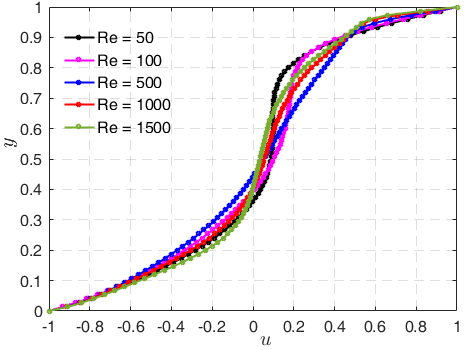}
  \caption{$u$-velocity along the vertical line through the geometric center of the domain.}
  \label{f5}
 \end{figure}

Based on the numerical solutions for $u$-velocity, Fig. \ref{f5} illustrates the variation of $u$-velocity along the vertical line through the geometric center of the rectangular domain at Reynolds numbers $Re=50, 100, 500, 1000$, and 1500. We can see that $u$-velocity increases from the bottom wall to the top wall of the rectangular domain.

 \begin{figure}[H]
  \includegraphics[scale=0.8]{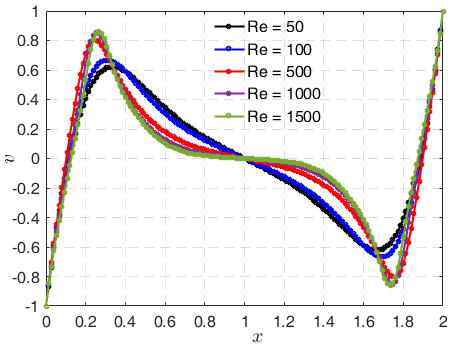}
  \caption{$v$-velocity along the horizontal line through the geometric center of the domain.}
  \label{f6}
 \end{figure}

Based on the numerical solutions for $v$-velocity, Fig. \ref{f6} illustrates the variation of $v$-velocity along the horizontal line through the geometric center of the rectangular domain at Reynolds numbers $Re=50, 100, 500, 1000$, and 1500. We can see that as Re is increased from 50 to 1500 the $v$-velocity profile looks more negative which is not the case for individual Reynolds numbers. We found that, $v$-velocity becomes almost similar in shape with increase Reynolds number. So, an oscillatory flow of $v$-velocity profiles has been observed in the domain.

 \begin{figure}[H]
  \includegraphics[scale=0.75]{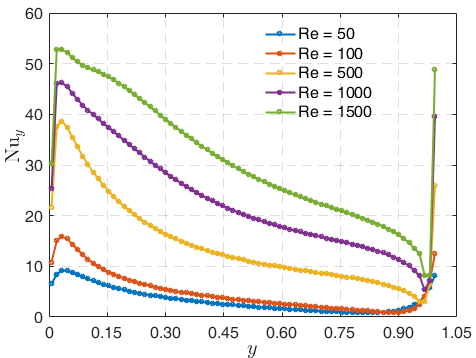}
  \caption{Local Nusselt number along the cold wall.}
  \label{f7}
 \end{figure}

Based on the numerical solutions for the local Nusselt number, Fig. \ref{f7} illustrates the variation of the local Nusselt number along the cold wall of the rectangular domain at Reynolds numbers $Re=50, 100, 500, 1000$, and 1500. We observed that, for each Reynolds number, the local Nusselt number decreases and thereby there is an increase in decay of heat.  

\begin{figure}[H]
  \includegraphics[scale=0.8]{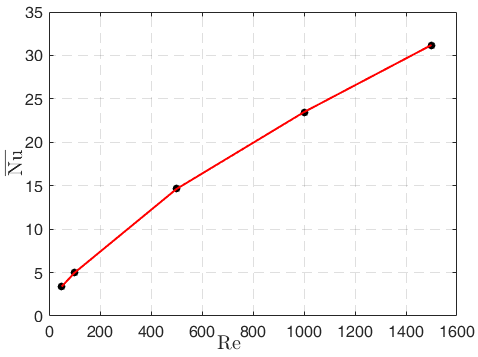}
  \caption{Average Nusselt number as a function of Reynolds number.}
  \label{f8}
 \end{figure}
 
\textit{Overall heat transfer}: Based on the numerical solutions for the average Nusselt number, Fig. \ref{f8} illustrates the variation of the average Nusselt number as a function of Reynolds number. We can see that, as $Re$ is increased from 50 to 1500 the average Nusselt number is also increased. We observe that overall heat increases as increase of Reynolds numbers.

\section{Conclusions}
\label{sec5.}

This paper presents a numerical study to examine the effects of moving walls on the fluid flow and heat transfer in a four-sided lid-driven rectangular domain insulated boundary conditions are imposed on horizontal side walls. The governing equations are solved using the SIMPLE finite volume method. We compared previously published work on special cases of the problem and found good agreement. We presented and discussed graphical results for various paramtric conditions. Heat transfer mechanisms and flow characteristics inside a rectangular domain are strongly dependent on moving wals and the Reynolds number. As the Reynolds numbers increase, so do both the local and average Nusselt number.

Numerical solutions for u-velocity illustrate the variation of $u$-velocity along the vertical line through the geometric center of the rectangular domain at Reynolds numbers $Re=50, 100, 500, 1000$, and 1500. We observed that $u$-velocity increases with increasing values of Reynolds from the bottom wall to the top wall of the rectangular domain. The numerical solutions for $v$-velocity illustrate the variation of $v$-velocity along the horizontal line through the geometric center of the rectangular domain at Reynolds numbers $Re=50, 100, 500, 1000$, and 1500. We can see that as $Re$ is increased from 50 to 1500 the $v$-velocity profile looks more negative which is not the case for individual Reynolds numbers. We found that, $v$-velocity becomes almost similar in shape with increase Reynolds number. So, an oscillatory flow of $v$-velocity profiles has been observed in the domain.
   
From streamlines plots for Reynolds numbers 50, 100, 1000, and 1500 of the rectangular domain, we observe, there is a primary cell with two secondary cells: the primary cell is at the center of the domain; however, the cell is not quite centered on the symmetry lines. The two secondary cells are weak and formed near the moving walls on both sides. While for Reynolds number 500, we observed that there are two secondary cells only. Both weaker cells are near the moving left and right walls. We also observe that streamlines become dense near the moving left and right walls. This happens due to the rises of the fluid along the right cold wall and sinks on the left hot wall due to forces generated by the moving fluid. Formation of two rotating cells at each side and a rotating cell at the center makes the heat transfer from left to right possible. The same phenomena have not been observed with one and two vertical-sided lid-driven cavities in the literature isotherms showed that, as Reynolds numbers increase the horizontal temperature gradient near the vertical walls decreases, because of which heat transfer decreases.
 
Numerical solutions for the local Nusselt number illustrates the variation of local Nusselt numbers along the cold wall of the rectangular domain at Reynolds numbers 50, 100, 500, 1000, and 1500. We observed that, for each Reynolds number, the local Nusselt number decreases and thereby there is an increase in decay of heat. The numerical solutions for the average Nusselt number illustrates the variation of average Nusselt numbers as a function of Reynolds number. We observe that the average Nusselt number, or overall heat, increases with increasing Reynolds numbers.

\mbox{}

\nomenclature{${p}'$}{pressure, $Pa$}
\nomenclature{$P$}{dimensionless pressure}
\nomenclature{$A$}{aspect ratio, $H/L$}
\nomenclature{$H$}{domain height, $\text{m}$}
\nomenclature{$L$}{domain length, $\text{m}$}

\nomenclature{$Re$}{Reynolds number, $V_pL/\nu$}
\nomenclature{$Pr$}{Prandtl number, $\nu/\alpha$}
\nomenclature{$Nu$}{local Nusselt number}

\nomenclature{$\overline{Nu}$}{average Nusselt number}
\nomenclature{$V_p$}{lid-driven plate velocity, $\text{m}/\text{s}$}

\nomenclature{$\alpha$}{thermal diffusivity, $\text{m}^2/\text{s}$}

\nomenclature{$\mu$}{dynamic viscosity, $\text{Kg}/\text{ms}$}
\nomenclature{$\nu$}{kinematic viscosity, $\text{m}^2/\text{s}$}
\nomenclature{$T$}{temperature, $\text{K}$}
\nomenclature{$\theta$}{dimensionless temperature}

\nomenclature{$t$}{non-dimensional time, $\text{s}$}
\nomenclature{${x}'$, ${y}'$}{cartesian coordinates}
\nomenclature{$x$, $y$}{dimensionless cartesian coordinates, ${x}'/L,\,{y}'/H$}

\nomenclature{$\Delta t$}{time spacing}
\nomenclature{$u$, $v$}{dimensionless velocities in $x$- and $y$ direction, $\text{m}/\text{s}$}

\nomenclature{${u}'$, ${v}'$}{velocities components in $x$, $y$ direction, $\text{m}/\text{s}$}

\nomenclature{$\dfrac{\partial}{\partial n}$}{differentiation along the normal to the boundary}

\nomenclature{$\rho$}{fluid density, $\text{kg}/\text{m}^3$}
\nomenclature{$\psi$}{stream function}

\nomenclature{$c$}{cold wall}
\nomenclature{$h$}{hot wall}
\nomenclature{$p$}{plate}
\nomenclature{$w$}{wall}

 \section*{Acknowledgments}

The first author acknowledge the support from research council, University of Delhi for providing research and development grand 2015--16 vide letter No. RC/2015-9677 to carry out this work.

\end{document}